\documentclass{aa} 
\usepackage{amsmath}
\usepackage{graphicx}
\usepackage{color}
\usepackage{txfonts}
\usepackage[authoryear]{natbib}

\begin{document}

\title{Observational evidence for a changing tilt of the accretion disk
       with respect to the orbital plane in Her X-1 over its 35 day cycle}

\author{D.~Klochkov \inst{1,2}, N.~Shakura \inst{2}, K.~Postnov\inst{2},
     R.~Staubert\inst{1}, J.~Wilms\inst{3,1}, N.~Ketsaris\inst{2}}

\offprints{klochkov@astro.uni-tuebingen.de}

\institute{
	Institut f\"ur Astronomie und Astrophysik, University of T\"ubingen,
	Sand 1, 72076 T\"ubingen, Germany
\and
	Sternberg Astronomical Institute, 119992, Moscow, Russia
\and
	Department of Physics, University of Warwick, Coventry, CV8 1GA, UK
}
\date{}
\authorrunning{Klochkov et al.}
\titlerunning{Disk inclination changes in Her X-1}

\abstract
   {}
   {
   Analysis and interpretation of the Her X-1 X-ray light curve obtained 
   with the ASM onbord RXTE over the period 1996 February to 2004 September.
   }
   {
   We construct the averaged X-ray light curves by means of adding up
   light curves corresponding to different 35\,day cycles. A numerical model
   is introduced to explain the properties of the averaged light curves.
   }
   {
   We report that the features found previously in the averaged X-ray lightcurve
   are confirmed by the new RXTE/ASM data. In particular, anomalous dips and 
   post-eclipse recoveries in two successive orbits in the short-on state are 
   found to be prominent and stable details of the light curve. We argue that a
   change of the tilt of the accretion disk over the 35\,d period is necessary to account 
   for these features and show that our numerical model can explain such a 
   behavior of the disk and reproduce the observed details of the light curve.
   }
   {}	   

\keywords{accretion, accretion disks--
                stars:binaries:general --
                X-rays: general  --
                X-rays: stars
               }
   
   \maketitle
%

\section{Introduction}

Her X-1/HZ Her is a close binary system with a 1.7\,d orbital period
consisting of an accretion powered 1.24\,s X-ray pulsar \citep{Giacconi73,
Tananbaum72} and an optical companion -- a main sequence star of spectral 
type A/F \citep{Crampton74}. The X-ray flux of the source shows a $\sim$35\,d
periodicity which is thought to be due to a counter-orbitally precessing 
tilted accretion disk around the neutron star \citep{GerBoy76}.

The main feature of the X-ray light curve is the alternation between {\em on} 
(high X-ray flux) and {\em off} (low X-ray flux) states. The 35\,d cycle
contains two {\em on} states -- the {\em main-on} ($\sim$7 orbital periods)
and the {\em short-on} ($\sim$5 orbital cycles) -- separated by $\sim$4-5 
orbital cycles. The maximum X-ray flux of the main-on is 
$\sim$5 times higher than that in the maximum of the short-on. The {\em off} states 
are usually explained by periodic obscuration of the X-ray source by the disk 
which most likely has a twisted form \citep{GerBoy76}. 

According to the generally accepted model 
\citep[see e.g.][ and references therein]{SchandlMeyer94,Kuster05}
the main-on state starts when the outer disk rim opens the line of sight to the 
source. Subsequently, at the end of the main-on state the inner part of the disk 
(that may be surrounded by a hot rarefied corona) covers the 
source from the observer. 
This picture seems to be confirmed by spectral analysis. The
opening of the X-ray source is accompanied by notable spectral changes including
evidence for the presence of a strong absorption, whereas the decrease in
X-ray intensity occurs more slowly and without appreciable spectral changes
(\citealt{Giacconi73, Kuster05}, see however \citealt{Oosterbroek00}).   

The transition from the {\em off} to the main-on state is usually called 
the {\em turn-on} of the source. An interesting fact is that turn-ons occur
near orbital phases $\phi_{\rm orb} \simeq$ 0.2 or $0.7$. This behavior has been 
explained by \citet{LevJer82} and \citet{Katz82}. These authors assumed that the disk 
undergoes a wobbling with half synodal period due to tidal torques. 
At $\phi_{\rm orb} \simeq 0.2$ and $0.7$ the angle between the disk plane 
and the observer's line of sight changes most rapidly. 

From the data described above it follows that the duration of the 35\,d cycle 
is a half-integer number of orbital cycles. Observations show that in most 
cases the duration of 35\,d cycles is 20, 20.5 or 21 orbital cycles 
\citep{Staubert83,Staubert06b}.  

Another interesting feature of the X-ray light curve is a sudden decrease in 
flux accompanied by significant spectral changes -- the {\em X-ray dips}. 
The dips were observed many times by different instruments
\citep[][ and references therein]{Shakura99}.
X-ray dips can be separated into three groups: {\em pre-eclipse dips}, which 
are observed in the first several orbits after X-ray turn-on, and march from 
a position close to the eclipse toward earlier orbital phase in successive
orbits; {\em anomalous dips}, which are observed at $\phi_{\rm orb} = 0.45 - 0.65$ 
and {\em post-eclipse recoveries}, which are occasionally observed as a short 
delay (up to a few hours) of the egress from the X-ray eclipse in the first 
orbit after turn-on.

\begin{figure}
\resizebox{\hsize}{!}{\includegraphics{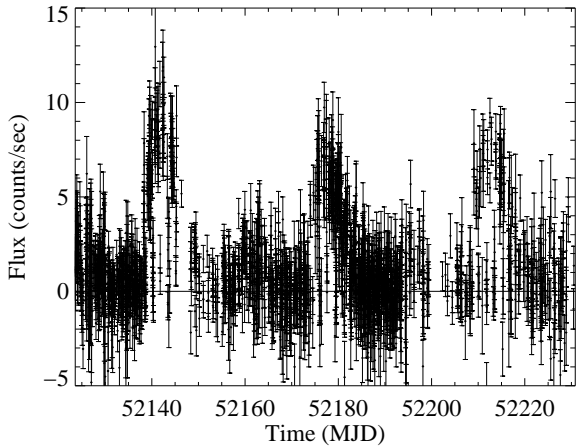}} 
\caption{\small Her X-1 ASM light curve sample.}
\label{asm}
\end{figure}

Several physical models have been proposed to explain the origin
and properties of the X-ray dips. For example, in many models 
\citep{CrosaBoynton80, Gorecki82, Leahy97, Schandl96} the dips 
are caused by thickening of the outer rim of the accretion disk  
due to interaction with the accretion stream; 
\citet{VrtilekHalpern85} suggested that the dips 
are produced by the accretion stream above and below the disk
\citep[see also][]{BochkarevKoritskaia89}.
However, these models made various assumptions
about the disk parameters (for example, in the original model 
by Crosa and Boynton the disk size was uncomfortably large 
to explain the X-ray dip marching behaviour).

Another model for X-ray dips was proposed by \citet{Shakura99} 
In this model pre-eclipse dips occur when the 
accretion stream crosses the observer's line of sight before entering the 
disk. This can happen only if the stream is non-coplanar to the systems orbital
plane. The reason for the stream to move out of the orbital plane 
is non-uniform X-ray heating of the optical stars atmosphere by the X-ray 
source, which produces a temperature gradient near the inner Lagrange point. 
The non-uniformity of the heating comes from the partial shadowing of the 
optical star surface by the accretion disk.
Furthermore, such a stream forces the outer parts of the 
accretion disk to be tilted with respect to the orbital plane. The tidal 
torques cause the disk to precess in the direction opposite to the orbital 
motion. Due to tidal torques and the dynamical action of the accretion stream 
the outer parts of the disk develop a notable wobbling (nutational) motion 
twice the synodal period. For an observer, the X-ray source can be screened 
by the outer parts of the disk for some time during the first orbit after the 
X-ray turn-on. This causes anomalous dips and post-eclipse recoveries.

The analysis of Her X-1 observations obtained with the {\em All Sky Monitor} (ASM)
on board of the {\em Rossi X-ray Timing Explorer} (RXTE) satellite \citep[see][]{Bradt93}
 over the period 1996 February to 2004 September allowed to improve the statistics 
of the averaged X-ray light curve of Her X-1 with respect to previous analyses 
\citep{Shakura98a,ScoLea99,StiBoy04}. 
In addition to well-known features 
(pre-eclipse dips and anomalous dips during the first orbit after X-ray 
turn-on) the averaged X-ray light curve shows that anomalous dips and 
post-eclipse recoveries are present for {\em two} successive orbits after the 
turn-on in the short-on state. These details have already been reported by 
\citet{Shakura98a}.
Now using about two
times more data we confirm these features to be stable details of the
light curve (they are also seen in one RXTE/PCA pointed observation of the source 
during the short-on state \citep{InamBaykal05}).

In this work we present averaged X-ray light curves of Her X-1 obtained from 
the analysis of ASM observations. Based on this data we modify
the model developed by \citet{Shakura99} in order to reproduce the details of the
averaged light curve which were left unexplained  previously.
We show that the accretion disk needs to change its tilt during the 35\,d
cycle from $\sim 20^\circ$ in the main-on to $\sim 4^\circ$ in the short-on.

\section{Processing and analysis of the X-ray light curve}

For the analysis of the X-ray light curve of Her X-1 we use data from the
ASM \citep{Levine96}. The archive contains X-ray flux 
measurements in the 2-12 keV band, averaged over $\sim$90 s. The monitoring 
began in February 1996 and continues up to date. The archive is
public and accessible on the Internet 
\footnote{http://xte.mit.edu/asmlc/srcs/herx1.html}.

X-ray data analyzed in this work contain some 50000 individual flux
measurements. They cover $\sim$90 35\,d cycles.
Preliminary processing of the X-ray light curve was carried out by
the method described by \citet{Shakura98a}. 

The goals of this processing are the reduction of the dispersion 
of the flux through rebinning, resulting in a smoothed light curve, and
the determination of the turn-on time of each individual 35\,day
cycle, with the aim to classify the cycles into two classes: turn-on 
around orbital phase $\sim 0.2$, and turn-on around orbital phase 
$\sim 0.7$ (they appear with about equal probability)

The smoothing of the light curves is achieved by combining appropriate
groups of data points into an average (following the procedure
described in \citet{Shakura98a}). The method used to determine the
turn-on time of the cycles without data gaps (we call them "good" cycles) 
is also similar to the one described by \citet{Shakura98a}. 
For cycles with data gaps longer than one day (we call them "bad" cycles)
a different method is used to determine the turn-on time. Each "good" 
cycle was fitted with an analytical function of the following form:
$$
f(t) =
A_0\frac{1}{1+\exp{((t-A_1)/A_2)}}\left( 1-\exp{\frac{t-A_3}{A_4}}\right),
$$
where $t$ is the time, $A_0,..,A_4$ are the fitting parameters. This
function consists of two smoothed ``steps'', the first representing
the beginning of a main-on state, the second representing the
end of the main-on. The analytical curves obtained from fitting "good"
cycles were used as a template library, assuming that each "bad" cycle can be
well described by one of these analytical curves. For a given "bad" cycle we find 
a corresponding curve from our library that fits it best in terms of
$\chi^2$, and we determine the turn-on time from this curve.

\begin{figure}
\resizebox{\hsize}{!}{\includegraphics{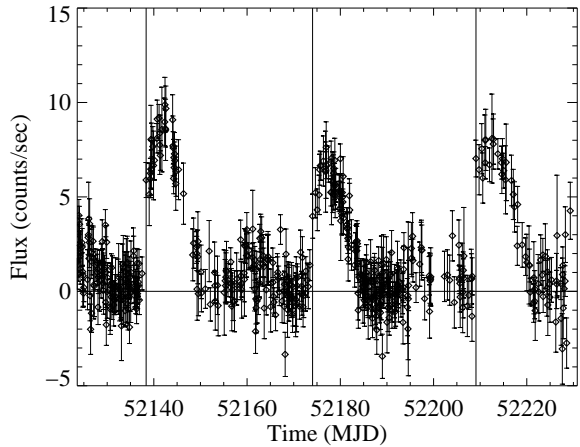}}
\caption{\footnotesize The same sample as in the Fig.~\ref{asm} after
processing. Vertical solid lines mark the time of X-ray 
turn-ons.}
\label{asmred}
\end{figure}

Figure~\ref{asm}  shows an example of the original X-ray
light curve. Figure~\ref{asmred} represents the same light curve
after processing.

\subsection{Averaged X-ray Light Curves}
 
Using the RXTE/ASM (with $\sim$5\,cts/s from Her~X-1 in the main-on)
details cannot be explored in individual 35 day cycles.
However, through superposition of many 35\,d light curves
common details (e.g. X-ray dips, post-eclipse recoveries) become recognizable.

This has been done e.g. by \citet{Shakura98a,ScoLea99,StiBoy04}. The ASM
archive has grown considerably, allowing to construct averaged light
curves with smaller dispersion than in previous works.

\begin{figure*}
\centering
\includegraphics[width=17cm]{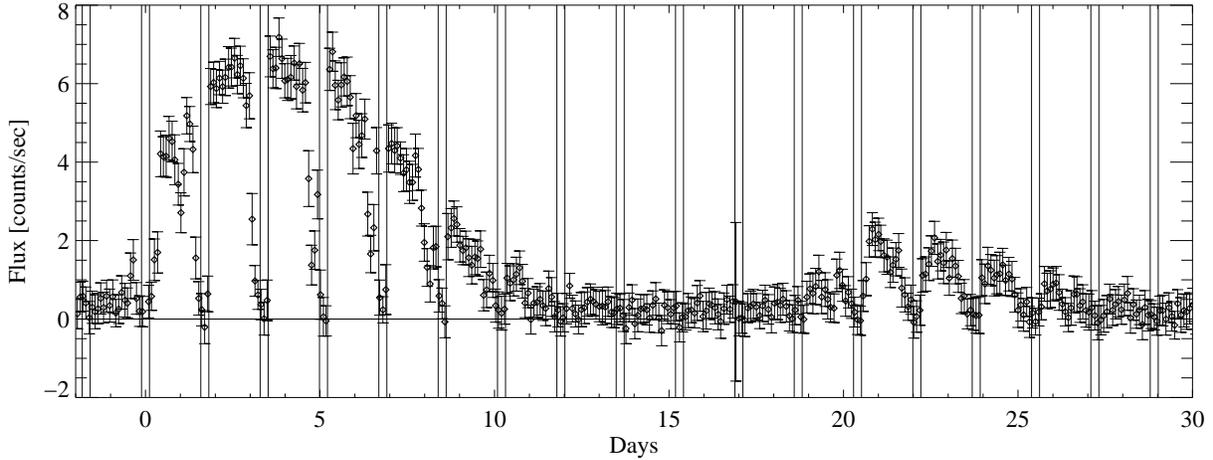}
\caption{\footnotesize Averaged X-ray light curves of \hbox{Her
X-1} corresponding to cycles with turn-ons near orbital phase
$\sim$0.2. Vertical lines show intervals of X-ray eclipses.}
\label{av25}
\end{figure*}

\begin{figure*}
\centering
\includegraphics[width=17cm]{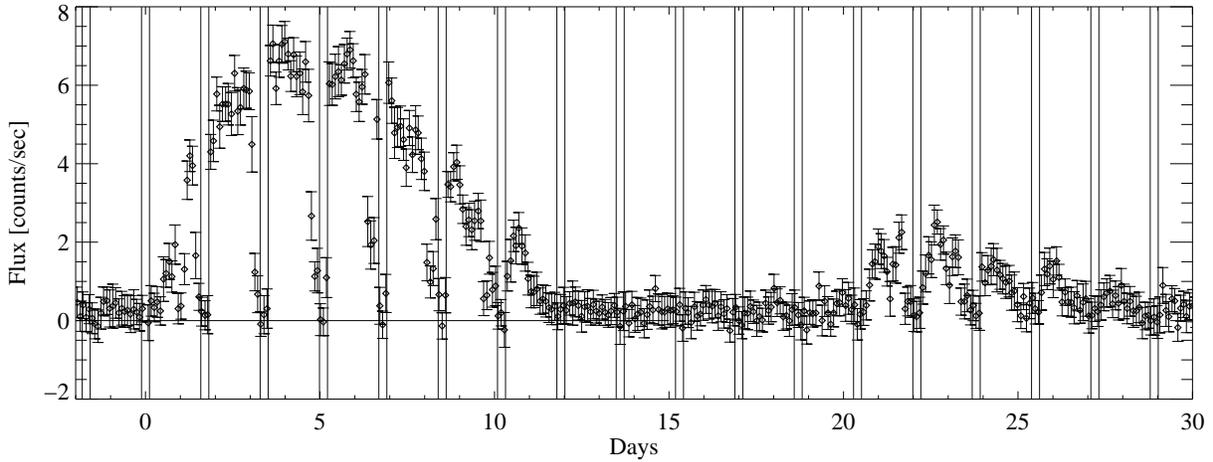}
\caption{\footnotesize The same as in the Fig.~\ref{av25} but for cycles
with turn-ons near the orbital phase $\sim$0.7.}
\label{av75}
\end{figure*}

As mentioned above, in most cases turn-ons occur near orbital phases 0.2 
and 0.7. Thus all 35\,d cycles have been divided
into two groups -- with the turn-on near $\phi_{\rm orb} = 0.2$ and
near $\phi_{\rm orb} = 0.7$. Inside each group the light curves were 
superposed and averaged, after shifting them in such a way that the
eclipses coincided.
The superposed light curves are shown in Figs.~\ref{av25} and \,\ref{av75}.

In the averaged light curve corresponding to the turn-on near 
$\phi_{\rm orb} = 0.7$ (Fig.~\ref{av75}) one can see that the flux
starts to increase near $\phi_{\rm orb} = 0.2$. But around 
$\phi_{\rm orb} = 0.5$ this increase is suppressed by a strong anomalous dip. 
This is due to the fact that some cycles classified to
turn-on near $\phi_{\rm orb} = 0.7$ have some increased flux near 
$\phi_{\rm orb} = 0.2$ (just before the anomalous dip), but
statistically insignificant to be recognized in the individual light curve.
Only when the increase of the flux before the anomalous dip is recognizable 
in the individual light curve we consider the turn-on as occurring near
$\phi_{\rm orb} = 0.2$. This shows some limitations of determining the 
turn-on times from RXTE/ASM data.

\begin{figure}
\resizebox{\hsize}{!}{\includegraphics{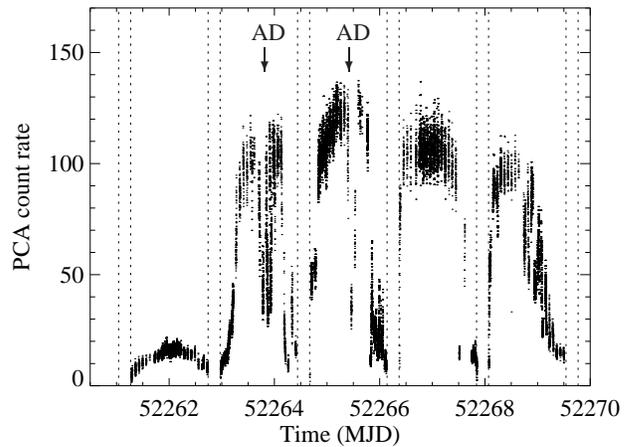}}
\put(-105,158){\vector(0,-1){9}} \put(-110,161){{\footnotesize AD}}
\put(-137,158){\vector(0,-1){9}} \put(-142,161){{\footnotesize AD}}
\caption{3-20 keV PCA light curve of Her X-1 during a short-on state.
Arrows show anomalous dips (AD) which are present on two successive 
orbits after the beginning of the short-on state}
\label{pcalc}
\end{figure}

In the short-on state complicated dip patterns can be observed. 
(Fig.~\ref{pcalc}). Anomalous dips near orbital phase $\phi_{\rm orb} = 0.5$ are 
present on two successive orbits after the beginning of the short-on state 
\citep[see also][]{Leahy00, Oosterbroek00, InamBaykal05}. 

\subsection{Statistics of individual 35\,d cycles}

The behavior of the duration of the 35\,d cycle can be studied using 
the so-called ($O-C$) diagram -- the difference between the observed 
and calculated times (assuming a constant period) of the turn-on as 
a function of time (or 35\,d cycle number). There is a 
strong correlation between ($O-C$) and the pulse period development,
both on long (years) and short (months) time scales 
\citep[see also][]{Staubert00,Staubert06a}. In addition, 
a lag is observed in the sense that structures in the curves tend to
appear first in the pulse period curve and then (about 35\,days later)
in the ($O-C$) diagram. This provides support to the model described 
here (see Discussion).
A comprehensive analysis of the behavior of the 35\,d cycle, pulse period 
development ($P$ and ${\dot P}$) and X-ray flux will be presented in a 
forthcoming paper by \citet{Staubert06b}. We will show that the
correlation to X-ray flux is comparatively weak. The statistics of the
duration of the 35\,day cycles is the following: about 40\% have a duration 
of $\sim 20.5$ binary cycles, the rest is equally distributed between cycles 
with $\sim 21.0$ and $\sim 20.0$ binary cycles (including a few cycles of 
duration $\sim 19.5$ binary cycles).

\section{Evidence from the X-ray light curve for a change in the tilt of the disk}

The fact that during the short-on state of Her X-1 more absorption dips 
appear on the X-ray light curve  (the anomalous dips and 
post-eclipse recoveries during two successive orbits of the 
averaged short-on as well as in an individual short-on observed by PCA:
Fig.~\ref{pcalc} and \citealt{InamBaykal05}) suggests that the  
angle $\epsilon$ between the disk and the line of sight remains close to zero 
during the first two orbits after the turn-on in the short-on state. This is 
in contrast to the main-on state, where such features are observed only during
one orbit after turn-on. This observation implies a generic asymmetry between
the beginning of the main-on and short-on states. 
This asymmetry can 
take place only if the disk tilt, $\theta$, changes with precession phase 
$\phi_{\rm pre}$. Indeed, the angle $\epsilon$ of the plane of the disk with 
the line of sight is 
\begin{equation}
\sin\epsilon = \cos i\cos\theta + \sin i\sin\theta\cos\phi_{\rm pre}\,.
\label{epsilon}
\end{equation}
where $i$ is the binary inclination angle. So with constant $i$ and 
$\theta$ ~~$|d\epsilon/d\phi_{\rm pre}|\sim|\sin\phi_{\rm pre}|$ 
and are the same for precession phases with $\epsilon=0$ 
(i.e. at the beginning of the {\em on}-states), although the line of 
sight to the X-ray source lies nearer to the plane of the disk during the
short-on state. As the binary inclination 
angle $i$ could not change significantly over the precession period, 
the only way to recover the asymmetry is to suppose that $\theta$ is a 
function of $\phi_{\rm pre}$.  

\begin{figure}
\resizebox{\hsize}{!}{\includegraphics{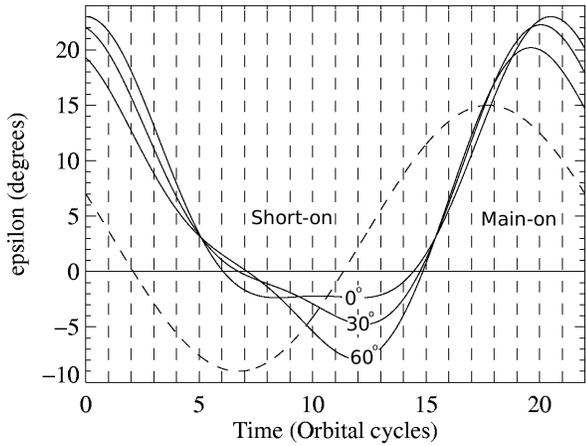}}
\caption{The effect of changing tilt of the disk on the angle $\epsilon$ 
between the line of sight and the outer plane of a flat disk (solid lines) 
for different phase delays $\Delta=0$, $30^\circ$ and $60^\circ$. The phase delay
causes the asymmetry in the short-on and main-on states. 
The larger $\Delta$, the more asymmetric is the changing of $\epsilon$. 
The dashed sine-like curve shows 
the changing of $\epsilon$ for the inner parts of the twisted disk.}
\label{f:delta_effect}
\end{figure}

The physical reason why this could be so is discussed below.  
The disk tilt is generally determined by the action of various torques 
acting on the disk. These include (1) the tidal torque from the optical star, (2) 
the dynamical torque caused by gas streams, (3) the viscous torque, (4)  
the torque caused by the radiation pressure and possibly (5) a disk wind torque. 
The joint action of these
torques determines the shape of tilted and twisted accretion disk (see, for 
example, calculations of the stationary disk shape for the disk wind model 
in Schandl \& Meyer 1994). However, if the mass transfer rate through the 
inner Lagrange point into the disk changes with the precession phase, so will
the tilt of the disk. 
For example, if free precession of the neutron star is ultimately 
responsible for the 35-day cycle in Her X-1
\citep{Brecher72,Truemper86,Shakura95,Shakura98b}, 
the conditions of  X-ray 
illumination of the optical stars atmosphere will periodically change with 
the precession phase. This in turn will lead to changes in the velocity 
components of the gas stream in the vicinity of the inner Lagrangian point and
hence in the matter supply rate to the accretion disk.
Recently, X-ray pulse profiles evolution with the 35\,d 
phase was successfully reproduced both in the main-on and short-on
states in the model of freely precessing neutron star with complex 
surface magnetic field structure \citep{ketsaris00,Wilms03}.
As a first approximation the periodic change of the tilt of the disk 
with precessional phase can be parameterized in the form

\begin{equation}
\theta=\theta_0[a+b\cos(\omega_{\rm pre}t+\Delta)]\,,
\label{theta_kachanie}
\end{equation}
where $\omega_{\rm pre}$ is the 35\,d precessional frequency,
$\Delta$ is the phase delay, and $a$ and $b$ are numerical
constants. The phase delay $\Delta$ is zero for the observer whose 
line of sight coincides with the extrema of the disk tilt $\theta$, 
which is quite improbable. So in general one expects $\Delta\ne 0$. 
The effect of the disk's tilt changing with the angle $\epsilon$ is shown in 
Fig.~\ref{f:delta_effect} for different phase delays $\Delta=0$, $30^\circ$ and
$60^\circ$. It is seen that in a fairly broad range of $\Delta$ the possibility 
appears for $\epsilon$ to be close to zero during several orbits in the
short-on state. 

The anomalous absorption dips and post-eclipse
recoveries can be produced by the wobbling motion of the outer disk. The
wobbling of the outer parts of the disk occurs twice during the orbital binary period
and is primarily caused by the tidal torques \citep{Katz82, LevJer82}.
The tidal torque vanishes when the companion crosses the node line of the
disk, so a wobbling appears as a sine-like component with one half
of the synodic orbital period on top of the mean disk tilt. 
This  explains, in addition to the occurrence of the turn-ons at the 
preferred 0.2 and 0.7 binary phases, the appearance of
anomalous dips during the first orbit after the turn-on both in
the main-on and in the short-on states (see Fig. 5 in \citealt{Shakura99}).

The appreciable tidal wobbling (with the amplitude of order of one degree) can
be experienced by the disk with the outer radius $r_{\rm d}\sim 0.3a$ 
($a$ is the binary separation). Tidal forces also cause precession of the disk
retrograde to the orbital motion. Acting alone, these forces would result 
in a precessional period somewhat shorter than observed.  
To explain the observed precessional period of the disk (about 20.5$P_{\rm orb}$, where
$P_{\rm orb}$ is the orbital period), 
braking torques must be added. For example, according to our qualitative accretion 
stream-disk coupling model \citep{Shakura99}, the motion of the outer
parts of the disk is driven by the joint action of tidal forces
and dynamical pressure induced by the impact of the gas stream on the
accretion disk. The dynamical action of the gas stream over one 
binary period is opposite to the tidal force so the streams can 
slow down the tidal precession to the observed value. As mentioned above, an important 
feature of our model is that the stream
is non-complanar with the orbital plane, which is due to a
complex shape of the shadow produced by the tilted and twisted
accretion disk on the atmosphere of the optical companion (HZ~Her). 
In addition to braking the precessional motion of the disk,
the stream exerts a torque tilting the disk, which
counterbalances the viscous torque and therefore limits the 
tilt of the disk \citep{BardeenPetterson75}.
So the outer regions of the accretion disk are normally kept tilted at some equilibrium angle
defined by the joint action of viscous and dynamical torques 
(for an alternative explanation of the disk tilt induced by the coronal wind
see \citealt{SchandlMeyer94}). Our model has allowed us to
successfully reproduce the phase behavior of pre-eclipse dips as the result of
regular screening of the X-ray source by the out-of-the-orbital-plane
stream \citep{Shakura99}.

It is important to note that in this model the gas stream impacts the tilted 
disk deeply inside its outer radius (see Fig. 2 in \citealt{Shakura99}) at 
distances $r_{\rm imp} \sim 0.1a$ from the neutron 
star. Assuming the standard accretion 
disk, the viscous time from the impact region to the disk center is around 10 
days, and scales as $r^{-3/2}$. However, the region of the disk beyond 
$r_{\rm imp}$, where is no matter supply (called the ``the stagnation zone'') 
and which mediates the angular momentum transfer outwards for accretion to 
proceed, can react to perturbations induced by the stream impact much faster 
than on the viscous time scale. Indeed, in a binary system, tidal-induced 
standing structures can appear in the outer zone of the accretion disk 
\citep[see e.g.][ and references therein]{Blondin00} and perturbations 
of angular momentum can propagate through this region with a velocity close 
to the sound speed (while the matter will accrete on the much slower viscous 
time scale!). Recent analysis of broad-band variability of SS 433 \citep{Rev05} 
suggests that such a picture is realized in the accretion disk in that source. 

The physical picture of the change of the disks tilt adopted here is as follows. 
The rate of mass transfer $\dot M$ supplied by HZ Her (which 
for a given moment can be different from the  
accretion rate of the neutron star because the viscous time scale for mass
transport through the disk both delays and smoothes the mass flow.)
changes periodically over 
the precession cycle in response, for example, to changing illumination of 
the optical star atmosphere  due to free precession of the neutron star. 
The streams action causes the outer disk to precess slower than it would 
do if only tidal torque was acting, and it also changes the outer 
disk's tilt on the dynamical time scale of about 10 days (see below), which 
is sufficient to explain periodic disk tilt variations over the precession 
cycle. The wobbling of the outer disk is mainly due to tidal forces 
(the dynamical action of the stream provides minor contribution). 
The viscous time scale in the ``stagnation zone'' is much longer, so the 
viscous torques cannot smooth out the external disk variations.    

\begin{figure}
\resizebox{\hsize}{!}{\includegraphics{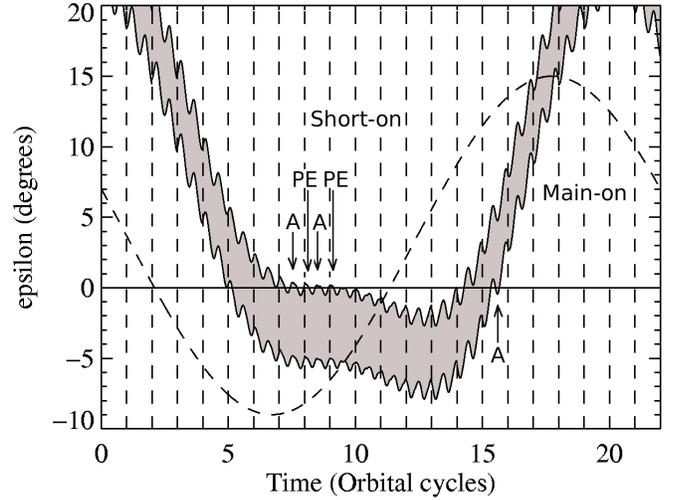}}
\caption{The angle $\epsilon$ between the direction from the center of
the neutron star to the observer and to the outer parts of the accretion 
disk with non-zero thickness. The dashed sine-like curve shows schematically the 
angle $\epsilon$ for the inner accretion disk regions which eclipses 
the X-ray source at the end of on-states. The vertical dashed lines mark 
centers of the binary eclipses.
Arrows mark anomalous dips (A) and post-eclipse recoveries (PE). The binary
system inclination $i$, parameters $\theta_0$ and $\Delta$ of the 
disk tilt change (Eq.\protect\ref{theta_kachanie})
and twisting angle of the disk (determining the location of the 
inner disk regions in this figure) are chosen such that to reproduce
the observed durations of the main-on and short-on states and the appearance
of two anomalous dips and post-eclipse recoveries in the beginning of the
short-on.}
\label{wobbl}
\end{figure}

In Appendix we present details of the calculation of the disk wobbling 
under the action of tidal and stream torques, and the result is shown in 
Fig.~\ref{wobbl} for the standard accretion disk semi-thickness $H/R=0.04$.
In these calculations, there are two characteristic radii: the outer 
radius of the disk $r_{\rm out}\sim 0.3a$ which determines 
the tidal wobbling amplitude, and the effective radius $r_{\rm eff}\sim 0.18a$ 
which determines the mean precession motion of the disk. 
The value of this effective radius is found from the requirement that 
the net precession period of the entire disk be equal to the observed 
value 20.5$P_{\rm orb}$ and the dynamical time scale $t_{\rm d}=M_{\rm d}/\dot M$ 
(which characterizes the dynamical action of the stream) be 10 days. 
This value for the dynamical time is chosen to be of order of the viscous 
time from the impact radius $\sim$0.1$a$ to meet (quasi) 
stationary conditions: at larger $t_{\rm d}$ the disk would rapidly empty out, 
at smaller $t_{\rm d}$ the matter would be stored in the disk. We emphasize again 
that the changing tilt of the outer parts of the disk (at $r_{\rm out}$) 
occurs very rapidly (on the sound velocity time scale!), while the 
precessional motion of the entire disk changes on much longer dynamical 
time scale $t_{\rm d}$ in response to the $\dot M$ variations.     
 
In Fig.~\ref{wobbl} we show the result of calculations (see Appendix) of 
the angle $\epsilon$ between the direction from the center of
the neutron star to the observer and to the outer parts of the accretion 
disk with non-zero thickness.
The complex shape of the disks wobbling is clearly seen. The sine-like dashed
curve shows schematically the angle $\epsilon$ for the inner accretion 
disk regions which eclipses the X-ray source at the end of on-states. 
The horizontal line is the observer's plane and the vertical
dashed lines mark centers of the binary eclipses. The main-on and
short-on states are indicated. 
The source is screened by the disk when the observer is in the magenta area
or between this area and the inner disk line. 
It is seen that the wobbling
effects can be responsible for the observed (several) anomalous
dips and post-eclipse recoveries at the beginning of the short-on
state. These additional dips become more pronounced with
the disk's tilt decreasing and possibly we see them as unusual
anomalous dips and post-eclipse recoveries at the end of the
short-on preceeding the anomalous low state observed by
BeppoSAX \citep{Oosterbroek00} and RXTE \citep{Still01}.

\section{Discussion}

For constructing the averaged X-ray light curves of \hbox{Her X-1} we used
the RXTE/ASM data covering $\sim$90 35\,d cycles. The large amount of data 
assures good statistics allowing to see some new details in the 
light curves. It is clearly seen that anomalous dips and post-eclipse
recoveries exist at two successive orbits after the turn-on in the short-on 
state. In our model these features are produced by the wobbling of the outer 
parts of the accretion disk which cover for some time the central X-ray 
source from the observer. To account for the observed behavior of the 
anomalous dips and post-eclipse recoveries we have to assume a change of 
the tilt of the outer parts of the disk with precessional phase. In the 
short-on state the tilt is three times smaller than in the main-on 
state. 

The model we used to calculate the wobbling of the outer parts of
the accretion disk in Her X-1 is highly idealized. Although 
detailed numerical calculations of precessing twisted accretion disks 
including viscous forces have been presented by several authors
\citep[e.g.][]{BardeenPetterson75,Petterson77},
only simplified models have been used to account for the wobbling motion
of the accretion disk \citep[see][ where only tidal wobbling is considered]{LevJer82}.
Thus, it is hardly now possible to check our model, where both tidal forces and
dynamical actions of the accretion stream are included, by numerical calculations. 
However, even
this simplified model explains most properties of the averaged X-ray
light curve of Her X-1: the duration of the main-on and short-on
states, the appearance and behavior of pre-eclipse dips
\citep{Shakura99} and of anomalous dips and post-eclipse
recoveries. 

In our calculations we ignored viscous torques exerted on the
outer regions of the disk. Taking them into account complicates
the model and can only be done introducing new assumptions about the
shape of the twisted and tilted accretion disk in Her X-1. We
expect that the effects of the viscosity will be the following. 
The viscosity always tends to reduce the tilt of the disk. It
counterbalances the dynamical action of the stream tending to
restore the tilt, so at any given time there is an equilibrium
tilt angle. Weakening the stream (for example, by decreasing the
accretion rate $\dot M$) would ultimately result in a smaller tilt
of the disk, which we believe is the physical reason for the appearance
of anomalous low states. 

In the above picture the time needed by the matter to travel from the 
stream impact region $r\sim 0.1a$ to the neutron star 
($\sim$10 days) is shorter than the viscous time of the disk at the effective 
radius $r_{\rm eff}\sim 0.18$ binary separation ($\sim$30 days). So the model 
predicts that the neutron star reacts to the accretion rate variations 
{\em earlier} than the precessing accretion disk. This can manifest itself, 
for example, in the neutron star period variations preceding changes in the 
accretion disk precession period. The analysis of long-term correlations 
in Her X-1 \citep{Staubert06b} indeed suggests such a delay between 
the turn-on behavior of the source on the O-C diagram and the pulse period 
variations and hence lends support to the model under discussion.    

\section{Conclusions}

The following main results have been obtained from analysing
and modeling RXTE/ASM X-ray light curves of Her~X-1.
\begin{enumerate}
\item The shape of the averaged X-ray light curves is determined more
accurately than was possible previously \citep[e.g. by ][]{Shakura98a}.
This has allowed us to confirm the significance and stability of the
features in the averaged short-on light curve which have been found previously: 
the anomalous dips in two successive orbits after the beginning
of the short-on state with a post-eclipse recovery after the first eclipse.
\item We argue that the changing of the tilt of the accretion disk with 
35\,d phase is necessary to account for the appearance of these features and
the observed duration of the main-on and short-on states.
\item We show that the joint action of tidal forces and the
dynamical interaction of the gas stream with the outer parts of the
accretion disk leading to its wobbling can be responsible for 
the observed behavior of the disk.
\end{enumerate}

\begin{acknowledgements}

In this research we used data obtained through the High Energy Astrophysics 
Science Archive Research Center Online Service, provided by the NASA/Goddard 
Space Flight Center.
We thank Prof. A. Santangelo for useful discussion.
The work was supported by the DFG grant Sta 173/31-2 and 
436 RUS 113/717/0-1 and the corresponding RBFR grant RFFI-NNIO-03-02-04003.
KP also acknowledges partial support through RFBR grant 03-02-16110. 
We thank the anonymous referee for constructive remarks.

\end{acknowledgements}

\bibliographystyle{aa}
\bibliography{refs}

\appendix
\section{Wobbling of the outer parts of the disk}

The new features discussed in this paper (anomalous dips
and post-eclipse recoveries in the first two successive orbits in
the short-on state) call for modification of the model proposed by 
\citet{Shakura99}. We argue that 
in order to explain these new features, the tilt of the outer
regions of the disk should regularly change over the 35\,d cycle. 
The physical reason for this can be a
decrease of the mass transfer rate $\dot M$ supplied by the optical component
as is expected in our model of a freely precessing neutron
star as the underlying clock mechanism for the 35\,d cycle in Her
X-1 \citep{Shakura99}.

To quantify this behavior, we calculated the net wobbling of the disk
produced by tidal forces and the accretion stream. In contrast to
\citet{SchandlMeyer94}, we ignored the viscous torques on the
twisted accretion disk (see discussion in Section 4). 

Calculations were done in the same way as
in \citet{Shakura99} allowing for the change of the tilt of the disk.

To account for the tidal torques, we approximated the outer disk
as a solid ring of radius $r_{\rm d}$ \citep[see discussion by ][]{Shakura99}. 
In the quadrupole approximation, the ring precesses around 
the orbital angular momentum with the frequency 
\begin{equation}
\omega_t \sim
\frac{3}{4}\frac{r_{\rm d}^{3/2}\cos\theta}{\sqrt{q+q^2}}\omega_{\rm b}\,,
\label{e:omega_tid}
\end{equation} 
where $q\equiv M_{\rm x}/M_{\rm o}\simeq 0.64$ is the binary mass ratio and 
$\omega_{\rm b}=2\pi/P_{\rm b}$ is the orbital binary frequency. 

We denote the precession angle of the disk by $\phi$ (counting
along the orbital rotation) and the tilt angle of the disk by $\theta$. 
The tidal wobbling of the disk on top of its slow precessional motion in 
the leading order in $\omega_{\rm t}/(2\omega_{\rm b}+\omega_{\rm t})\ll 1$ can be described as
\citep{LevJer82}
\begin{equation}
\frac{d\phi}{dt} 
\simeq -\omega_{\rm t}\left[1+\frac{2\omega_{\rm b}}{2\omega_{\rm b}+\omega_{\rm t}}\cos2(\phi_{\rm b}-\phi_{\rm pre})\right]\,, 
\end{equation}
\begin{equation}
\frac{d\theta}{dt}
\simeq
\omega_{\rm t}\sin\theta\frac{2\omega_b}{2\omega_{\rm b}+\omega_{\rm t}}\sin2(\phi_{\rm b}-\phi_{\rm pre})\,.
\end{equation}
where $\phi_{\rm b}$ is the orbital phase and $\phi_{\rm pre}$ is the precessional phase.

Let us now discuss the dynamical action of the stream in more detail. The
angular momentum of the outer parts of the disk ${\bf K}_{\rm d}$ are changed by 
the stream according to the equation
\begin{equation}
\frac{d{\bf K}_{\rm d}}{dt}=\dot M[{\bf r}\times{\bf v}]\,,
\label{dK/dt}
\end{equation}
where ${\bf r}$ and ${\bf v}$ are the distance from center of the disk
to the impact point of the stream and the gas velocity at the
impact point, respectively. They depend on the binary phase. In
our model $\dot M$ is also a function of the orbital phase
because the matter supply through the inner Lagrangian point
depends on its position relative to the shadow produced by the tilted and
twisted disk in the atmosphere of the optical star.

We will use the right-hand Cartesian reference frame rigidly
connected to the accretion disk and consider it as non-rotating
(this assumption is justified by the relatively slow precessional motion of
the disk). The origin of the frame is at the center of the neutron star, the
$z$-axis is normal to the orbital plane and the $x$-axis lies in
the orbital plane and is directed such that the disks 
angular momentum vector always lies in the $z-x$ plane (see
Fig.~\ref{ref_frame}):

\begin{figure}
\resizebox{\hsize}{!}{\includegraphics{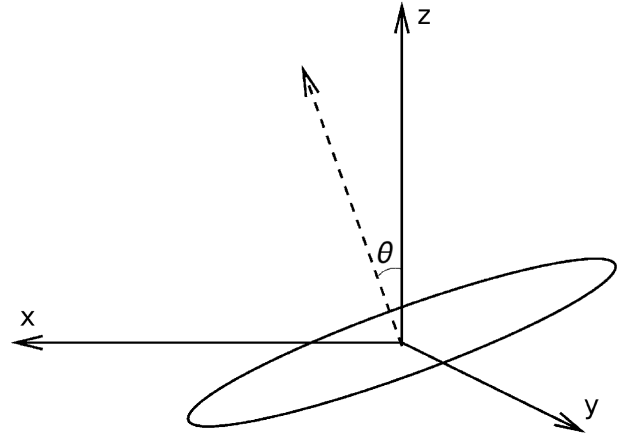}}
\caption{The reference frame connected to the disk. The origin of the frame is
  at the center of the neutron star. The vector of the angular momentum 
  ${\bf K}$ of the disk always lies in the $z-x$ plane.}
\label{ref_frame}
\end{figure}

\begin{equation}
{\bf K}_{\rm d}=(K_{\rm d}\sin\theta, 0, K_{\rm d}\cos\theta)\,.
\end{equation}
Then the change of the disk momentum $d{\bf K}_{\rm d}=(d{\bf K}_{\rm x},
d{\bf K}_{\rm y}, d{\bf K}_{\rm z})$ can be written through the change in its components
as
\begin{equation*}
dK_{\rm d}=dK_{\rm x}\sin\theta+dK_{\rm z}\cos\theta\,, 
\end{equation*}
\begin{equation}
K_{\rm d}\theta=dK_{\rm z}\cos\theta-dK_{\rm z}\sin\theta\,, \label{dK} 
\end{equation}
\begin{equation*}
K_{\rm d}\sin\theta d\phi=dK_{\rm y}\,. 
\end{equation*} 
The time evolution of the disk angular momentum components 
 (Eq.(\ref{dK/dt})) in this frame now can be written in the form:
\begin{equation*}
\frac{dK_{\rm x}}{dt}=\dot M v_{\rm rel} a (yv_{\rm z}-zv_{\rm y})\,, 
\end{equation*}
\begin{equation}
\frac{dK_{\rm y}}{dt}=\dot M v_{\rm rel} a (zv_{\rm x}-xv_{\rm z})\,, \label{dK/dt:comp}
\end{equation}
\begin{equation*}
\frac{dK_{\rm z}}{dt}=\dot M v_{\rm rel} a (xv_{\rm y}-yv_{\rm x})\,, 
\end{equation*}
where $x, y, z$  are dimensionless coordinates of the stream impact point 
(in units of binary separation $a$) and $v_{\rm x}, v_{\rm y}, v_{\rm z}$ are the 
components of the velocity of the stream at the impact 
point (in units of relative velocity of binary components $v_{\rm rel}$). 
For computing ballistic trajectories of the gas stream from the
Lagrangian point we used the primed right-hand reference frame $x',y',z'$ 
rigidly rotating with the binary system. The origin of the frame is at the 
center of the neutron star, the $z'$-axis is normal to the orbital 
plane and the $x'$-axis is pointed to the center of the normal component.  
The transition from that frame to the one of the precessing disk is 
made by usual coordinate transformation:

\begin{equation*}
x=x'\cos\Phi-y'\sin\Phi\,,
\end{equation*}
\begin{equation}
y=x'\sin\Phi+y'\cos\Phi\,, \label{trans}
\end{equation}
\begin{equation*}
z=z'\,,     
\end{equation*}
and for velocities:
\begin{equation*}
v_{\rm x}=v_{\rm x}'\cos\Phi-v_{\rm y}'\sin\Phi-\sqrt{(x')^2+(y')^2}\sin\Phi \,,
\end{equation*}
\begin{equation}
v_{\rm y}=v_{\rm x}'\sin\Phi+v_{\rm y}'\cos\Phi+\sqrt{(x')^2+(y')^2}\cos\Phi\,, \label{v_trans}
\end{equation}
\begin{equation*}
v_{\rm z}=v_{\rm z}'\,,  
\end{equation*}
where $\Phi$ is the angle between the coordinate axes $x$ and $x'$.

It is convenient to rewrite all equations in a dimensionless
form. The angular momentum of the disk can be written as
\begin{equation}
|{\bf K}_{\rm d}|=\gamma M_{\rm d}\omega_{\rm k}r_{\rm eff}^2\,,   \label{K_d}
\end{equation}
where $r_{\rm d}$ is the effective disk radius (see Section 3), $M_{\rm d}$ is the 
mass of the disk and
$\omega_{\rm k}=\sqrt{GM_{\rm x}/r_{\rm eff}^3}$ is the Keplerian frequency at the
effective radius and $\gamma$ is a numerical coefficient accounting for
the surface density distribution $\Sigma(r)$; for example, in the
standard Shakura-Sunyaev accretion disk \citep{ShakuraSunyaev73}
$\Sigma\propto r^{-3/4}$ and $\gamma=5/7$.

Now, from equations (\ref{dK}) and (\ref{dK/dt:comp}))  and making use of
the Kepler's 3rd law, we arrive at the following equations for the
change of angles $\phi$ and $\theta$ with the synodic angle
$\tau=t\omega_{\rm s}$ ($\omega_{\rm s}=\omega_{\rm pre}+\omega_{\rm b}$):

\begin{equation*}
\sin\theta\frac{d\phi}{d\tau}=\frac{\dot M}{\gamma
M_{\rm d}\omega_{\rm s}}\sqrt{\frac{M_{\rm x}+M_{\rm o}}{M_{\rm x}}}\sqrt{\frac{a}{r_{\rm d}}}[zv_{\rm x}-xv_{\rm z}]\,,
\label{dphi/dtau}
\end{equation*}
\begin{equation}
\frac{d\theta}{d\tau}=\frac{\dot M}{\gamma
M_{\rm d}\omega_{\rm s}}\sqrt{\frac{M_{\rm x}+M_{\rm o}}{M_{\rm x}}}\sqrt{\frac{a}{r_{\rm d}}} 
[\cos\theta(yv_{\rm z}-zv_{\rm y})-
\end{equation}
\begin{equation*}
\qquad -\sin\theta(xv_{\rm y}-yv_{\rm x})]\,.\label{dtheta/dtau}
\end{equation*}
Using these equations, we calculated the change in angles $\phi$ and $\theta$
($\Delta\phi$ and $\Delta\theta$) step by step and sum them up to
find their behavior over one precessional cycle. The initial components of 
the velocities $v_{\rm x}', v_{\rm y}', v_{\rm z}'$ were chosen as a function of the synodic 
phase $\Phi$ in the way described in detail by \citet{Shakura99}.

The free parameters are the mass of the accretion disk $M_{\rm d}$, 
the maximal disk tilt $\theta_{\rm 0}$ and its change, 
the disk semi-thickness $h/r_{\rm d}$ at the outer boundary and the 
binary inclination angle $i$. We fixed the angle $i=88\fdg 6$ and 
the disks semi-thickness $h/r_{\rm d}=0.04$ (corresponding to a disk opening 
angle of about $5^\circ$). The disks tilt was allowed to change with 
precessional phase according to the Eq.~(\ref{theta_kachanie}).

The angle $\theta_0=20^\circ$ and numerical coefficients ($a=0.6$, $b=0.4$ and 
$\Delta=15^\circ$) in Eq. (\ref{theta_kachanie})
were chosen to reproduce the observations. Now we can calculate the 
angle $\epsilon$ between the line of sight and the outer parts of the disk 
making use of Eq. (\ref{epsilon}).

In contrast to the case with constant tilt $\theta$ considered earlier 
by \citet{Shakura99}, the change of $\epsilon$ with precessional phase
is determined also by the change of the tilt $\theta(\phi_{\rm pre})$ [Eq.
(\ref{theta_kachanie})]. The effect of the phase delay $\Delta$ is
illustrated by Fig.~\ref{f:delta_effect}: the larger $\Delta$ is, the
more asymmetric is the changing of $\epsilon$
The results are shown in Fig.~\ref{wobbl}.

\end{document}